\documentclass[10pt,conference]{IEEEtran}
\usepackage{graphicx} 
\usepackage[utf8]{inputenc}
\usepackage{amsmath,amsfonts,amsthm}
\usepackage{algorithm}
\usepackage{algorithmic}
\usepackage{xcolor}
\usepackage{xspace}
\usepackage{xurl}
\usepackage{enumitem}
\usepackage{pifont}
\usepackage{multirow}
\usepackage{multicol}
\usepackage{makecell}
\usepackage[hidelinks]{hyperref}
\usepackage{tcolorbox}
\usepackage{subcaption}
\usepackage{caption}
\usepackage{textcomp}
\usepackage{listings}
\usepackage{balance}
\usepackage{fancyhdr}
\usepackage{balance}

\newcommand{\ddminlocalize}{\textsc{DDMIN-LOC}\xspace}
\newcommand{\ddmin}{\textsc{DDMIN}\xspace}

\title{Extending Delta Debugging Minimization for Spectrum-Based Fault Localization}

\author{\IEEEauthorblockN{Charaka Geethal Kapugama} \IEEEauthorblockA{\textit{Department of Computer Science} \\ \textit{Faculty of Science} \\ \textit{University of Ruhuna} \\ Sri Lanka \\ charaka@dcs.ruh.ac.lk}}

\date{November 2025}

\begin{document}

\maketitle

\thispagestyle{fancy}
\fancyhf{} 
\fancyfoot[l]{\small 979-8-3315-9397-1/26/\$31.00~\copyright~2026 IEEE}

\renewcommand{\headrulewidth}{0pt}

\begin{abstract}
    This paper introduces \ddminlocalize, a technique that combines Delta Debugging Minimization (\ddmin) with Spectrum-Based Fault Localization (SBFL). It can be applied to programs taking string inputs, even when only a single failure-inducing input is available. \ddmin is an algorithm that systematically explores the \emph{minimal failure-inducing input} that exposes a bug, given an initial failing input. However, it does not provide information about the faulty statements responsible for the failure. \ddminlocalize addresses this limitation by collecting the passing and failing inputs generated during the \ddmin process and computing suspiciousness scores for program statements and predicates using SBFL algorithms. These scores are then combined to rank statements according to their likelihood of being faulty. \ddminlocalize requires only one failing input of the buggy program, although it can be applied only to programs that take string inputs.  
    \ddminlocalize was evaluated on 136 programs selected from the QuixBugs and Codeflaws benchmarks using the SBFL algorithms Tarantula, Ochiai, GenProg, Jaccard and DStar.
    Experimental results show that \ddminlocalize performs best with Jaccard: in most subjects, fewer than 20\%  executable lines need to be examined to locate the faulty statements. Moreover,  in most subjects, faulty statements are ranked within the top 3 positions in all the generated test suites derived from different failing inputs.
\end{abstract}

\vspace{1em}

\section{Introduction}

The growing complexity of software systems has made manual debugging increasingly challenging. Nevertheless, software debugging remains essential for ensuring that software systems are fault free.  To address this challenge, various \emph{Automated debugging techniques}~\cite{parnin2011automated} are being introduced. Software industries are increasingly adopting these automated debugging techniques.

\emph{Spectrum-Based Fault Localization} (SBFL)~\cite{abreu2007accuracy} is one of the popular automated debugging techniques. These techniques take as input a test suite containing both passing and failing test cases, where the failing test cases expose the faulty behaviours of the system under test. SBFL techniques are based on the key assumption that faulty lines are executed more frequently by failing test cases than by passing test cases. SBFL techniques can be applied at the statement level and the predicate level~\cite{zhang2006locating}.  Tarantula~\cite{jones2005empirical}, Ochiai~\cite{abreu2007accuracy} and DStar~\cite{Dstar} are examples of SBFL algorithms. 

Zeller et al. introduced the concept of \emph{Delta Debugging}~\cite{zeller2002simplifying}. Their work presents two algorithms: \emph{Delta Debugging Minimization} (\ddmin), which finds the minimal failure inducing input from a given failing input, and \emph{Delta Debugging(DD)}, which isolates the minimal failure inducing difference between a passing and failing test case. \ddmin employs the \emph{binary search algorithm}~\cite{nowak2008generalized}. During the process of finding the minimal failure inducing input, \ddmin generates a set of intermediate passing and failing inputs. 

As suggested by Zeller et al.~\cite{zeller2002simplifying}, given a faulty program and a compiler, \ddmin can isolate the code segment that causes the compiler to crash, thereby helping to localize the fault. However, this technique cannot be applied to \emph{semantic bugs} or \emph{functional bugs}~\cite{tan2014bug}, in which program crashes are not typically observed. Furthermore, the minimal failing input of a semantic bug is inadequate to locate the faulty code segments that cause the bug. This work mainly focuses on addressing this issue by combining \emph{Delta Debugging Minimization} (\ddmin) with Spectrum-Based Fault Localization (SBFL). 

Since \ddmin employs a binary search strategy, the intermediate test inputs it generates are not produced randomly. The work \textsc{grammar2fix}~\cite{grammar2fix} shows that these test inputs can be used to approximate the condition under which a bug is exposed (i.e., \emph{the failure condition}). Moreover, \ddmin can serve as a technique for generating more test inputs from a single failure inducing input. Inspired by these concepts, \ddminlocalize leverages the test inputs generated by \ddmin together with spectrum-based fault localization.  

\ddminlocalize, presented by this paper, takes a failure-inducing input of the system under test. \ddmin is then applied to the failing input, and the passing and failing test cases generated during the process are collected. Using an SBFL algorithm, the suspiciousness scores for program statements and predicates are computed. These scores are then combined to calculate a hybrid suspiciousness score for each program statement. Based on these scores, the program statements are ranked according to their likelihood of being faulty. \ddminlocalize can be applied to programs taking string inputs. 

Several experiments were conducted using 136 programs from the Codeflaws~\cite{Tancodeflaws} and QuixBugs~\cite{lin2017quixbugs} benchmarks, which contain different types of real-world functional bugs. The SBFL algorithms-Tarantula~\cite{jones2005empirical}, Ochiai~\cite{abreu2007accuracy}, GenProg~\cite{GenProg}, Jaccard~\cite{chen2002pinpoint} and DStar~\cite{Dstar},- were tested with \ddminlocalize in these experiments. The experimental results demonstrate that:
\begin{enumerate}
    \item \ddminlocalize effectively localizes faulty statements when the Jaccard score is used with the hybrid approach. In this setting, a developer would typically need to examine less than 20\% of the executable lines, and the faulty statement is ranked within the top 3 positions. 
    \item The predicate-based method is unstable for certain types of bugs.
    \item The statement-based method exhibits the poorest performance across all the SBFL algorithms. 
\end{enumerate}

In summary, the \emph{main contributions} of this work are as follows. 
\begin{enumerate}
    \item This work introduces a technique for localizing software bugs using a single input that exposes the bug, i.e., a failing input. 
    \item It extends Delta Debugging Minimization (DDMIN)~\cite{zeller2002simplifying} for spectrum-based fault localization.
    \item A set of experiments is conducted on 136 program subjects containing diverse functional bugs to demonstrate that \ddminlocalize is effective in fault localization.
\end{enumerate}

\textbf{Reproducibility.} To facilitate reproducibility, the implementation of \ddminlocalize, the collected experimental data and the scripts have been made available at:

\textcolor{blue}{\url{https://github.com/charakageethal/ddmin_loc}}
\section{Methodology}

\lstset{
frame=single,
stepnumber=1,
 numbers=left,           
  xleftmargin=0.5cm,
  xrightmargin=0.15cm,
 breakatwhitespace=false,
  breaklines=true,
firstnumber=1,
 numberstyle=\scriptsize,
 basicstyle=\ttfamily\small,
tabsize=1,
}

Algorithm~\ref{alg:ddmin_loc} describes \ddminlocalize. First, \ddminlocalize takes as input a failing input $f$ for the given buggy program $\mathcal{P}$ and applies the Minimizing Delta-Debugging (\ddmin) algorithm~\cite{zeller2002simplifying} to $f$. Throughout the search for the minimal failing input $f_{\min}$, \ddmin systematically generates a set of passing inputs $T_p$ and a set of failing inputs $T_f$ for the given buggy program.  The statements and predicates of the buggy program executed by these test inputs ($T_p \cup T_f$) are recorded, and the suspiciousness scores for all statements and predicates are subsequently computed from this data (Line 4). Finally, the program statements are ranked according to their suspiciousness scores, reflecting their estimated likelihood of faultiness (Line 5). 

\begin{algorithm}
    \small
    \renewcommand{\algorithmicrequire}{\textbf{Input:}}
    \renewcommand{\algorithmicensure}{\textbf{Output:}}
    \newcommand{\algorithmicbreak}{\textbf{break}}
    \newcommand{\BREAK}{\STATE \algorithmicbreak}
    \caption{\textsc{DDMIN-LOC}}
    \label{alg:ddmin_loc}
    \begin{algorithmic}[1]
    \REQUIRE Buggy Program : $\mathcal{P}$
    \REQUIRE Initial Failing Input : $f$
    \STATE Let $T_p$ be the set of passing test inputs generated by \ddmin
    \STATE Let $T_f$ be the set of failing test inputs generated by \ddmin
    \STATE $T_p,T_f \gets \ddmin(f)$
    \STATE $S_E \gets \textsc{calculate\_suspiciousness\_scores}(T_p,T_f,\mathcal{P})$
    \STATE $E_R \gets \textsc{rank\_program\_statements}(S_E)$
    \RETURN $E_R$
    \end{algorithmic}
\end{algorithm}

\subsection{Spectrum-Based Fault Localization Algorithms}

\textbf{Spectrum-Based Fault Localization~(SBFL)} is founded on the intuition that program elements executed by many failing test cases and few passing test cases are likely to be faulty~\cite{abreu2007accuracy}. Many automated program repair techniques employ SBFL techniques to identify program statements that are likely to be faulty (E.g., GenProg~\cite{GenProg}, Angelix~\cite{mechtaev2016angelix}). Based on the execution frequencies of passing and failing test cases, SBFL techniques assign suspiciousness scores to program elements (statements or predicates). 

This paper considers five popular SBFL techniques, as described in Table~\ref{tab:fault_localize_alg}, for use with \ddminlocalize in the \textsc{calculate\_suspiciousness\_scores} function (Algorithm~\ref{alg:ddmin_loc}). In the following, let \textit{totalfailed} be the total number of failing test cases, and $\textit{failed}(s)$ be the number failing test cases executed statement $s$ (Similarly for \textit{totalpassed} and $\textit{passed}(s)$.

\begin{table}[htbp]
    \centering
    \begin{tabular}{ll}
        \hline
        \textbf{Name} & \textbf{Formula} \\
        \hline
       Tarantula~\cite{jones2005empirical}  & $S(s)=\frac{\textit{failed}(s)/\textit{totalfailed}}{\textit{failed}(s)/\textit{totalfailed}+\textit{passed}(s)/\textit{totalpassed}}$ \\[5pt]
       Ochiai~\cite{abreu2007accuracy} & $S(s)=\frac{\textit{failed}(s)}{\sqrt{\textit{totalfailed} \times (\textit{failed}(s)+\textit{passed}(s))}}$  \\[5pt]
        GenProg~\cite{GenProg} & $ S(s)=\begin{cases}
                    0 & \textit{failed}(s)=0 \\
                    1.0 &\textit{passed}(s)=0 \wedge\textit{failed}(s) > 0 \\
                    0.1 & \textit{Otherwise}
        \end{cases}$ \\ [5pt]
        Jaccard~\cite{chen2002pinpoint} & $S(s)= \frac{\textit{failed}(s)}{\textit{execute}(s)+(\textit{totalfailed}-\textit{failed}(s))}$ \\ [5pt]
         DStar~\cite{Dstar} & $S(s)=\frac{\textit{failed}(s)^{*}}{\textit{passed}(s)-(\textit{totalfailed}-\textit{failed}(s))}$\\
         \hline
    \end{tabular}
    
    In Dstar, the exponent(*) was set to 2 as reported in the prior study~\cite{PearsonICSE}.
    \caption{Spectrum-based fault localization algorithms used with \ddminlocalize}
    \label{tab:fault_localize_alg}
\end{table}

Tarantula~\cite{jones2005empirical} and Ochiai~\cite{abreu2007accuracy} are widely used SBFL techniques~\cite{gazzola2017automatic}. These two techniques have often served as baselines for evaluating the effectiveness of newly proposed SBFL methods~\cite{PearsonICSE}\cite{jiang2019combining}\cite{zhang2019empirical}. Therefore, these algorithms were included in this work. GenProg~\cite{GenProg} score and Jaccard score~\cite{chen2002pinpoint} were also considered in this work, as they are employed in automated program repair tools. In addition, DStar~\cite{Dstar} was included due to its demonstrated effectiveness in recent studies on fault localization.


\subsection{Fault Localization Based on Different Program Elements}\label{meth:prog_element}

SBFL techniques primarily focus on  evaluating the suspiciousness of \emph{program statements}. However, several studies have instead examined the suspiciousness of \emph{predicates}~\cite{jiang2019combining}\cite{liblit2005scalable}\cite{arumuga2007statistical}. According to the findings of \cite{liblit2005scalable} and \cite{jiang2019combining}, all conditional statements are treated as predicates, and using such predicates improves the accuracy of fault localization. Some studies compute a hybrid score that combines the suspiciousness scores of a statement and its associated predicates~\cite{xuan2014learning}\cite{zhang2019empirical}. 

This paper evaluates the fault localization effectiveness of the \emph{statement-based}, \emph{predicate-based} and \emph{hybrid} SBFL methods separately. In the predicate-based method, after calculating the suspiciousness scores of the predicates in a program, these scores are mapped back to their corresponding program statements. In the hybrid method, the hybrid score of a statement is calculated using the following equation.
\begin{equation}\label{eqn:hybrid}
\begin{split}
    \textit{HybridScore}(s)=\alpha\times \max(\textit{AssociatedPredicateScores}(s)) \\ +(1-\alpha)\times\textit{StatementScore}(s)
\end{split}
\end{equation}

This equation was inspired by the work of Xuan et al.~\cite{xuan2014learning}. The value of $\alpha$ determines the contributions from both predicate-based and statement-based approaches. 

\subsection{Example}
\begin{lstlisting}[language=Python, caption=Example Buggy Program,captionpos=b,label={lst:buggy_prog}][H]
def count_As_and_Es(word):
    count=0
    for w in word:
        if (w in ['a','d']): # buggy
            count+=1
    return count
\end{lstlisting}
The program in Listing~\ref{lst:buggy_prog} is intended to count the occurrences of the letters `a' and `e' in a given word. However, due to a mistake in Line 4, 'e's are not counted, resulting in a functional bug. For the word ``accurate", the total number of `a's and `e's is 3. Due to the aforementioned bug, this program returns 2; therefore, ``accurate" is a failing input. 

If \ddmin is applied to ``accurate", the sets of passing ($T_p$) and failing ($T_f$) inputs are as follows. 
\begin{itemize}
    \item $T_f=\{\text{``accurate",``rate",``te",``e"}\}$
    \item $T_p=\{\text{``accu",``ra",``t"}\}$
\end{itemize}

The suspiciousness scores for each statement and each predicate are calculated using an SBFL algorithm (Table~\ref{tab:fault_localize_alg}), taking $T_f$,$T_p$ and the program (Listing~\ref{lst:buggy_prog}) as inputs. The predicates in  Listing~\ref{lst:buggy_prog} are $\texttt{w} \in \texttt{word}$, $\texttt{w} \in [\texttt{'a','d'}]$ and $\texttt{w} \notin [\texttt{'a','d'}]$. These scores are then combined using Equation~\ref{eqn:hybrid}, and the statements are ranked according to their hybrid suspiciousness scores. 

\emph{Runtime analysis:} The worst-case runtime complexity of \ddmin is $O(n^2)$, where $n$ is the size of the failing input. This complexity can be reduced by caching previously executed test cases~\cite{zeller2002simplifying}, a technique adopted in this work. Moreover, the test cases generated by \ddmin during the process of exploring a minimal failure inducing input should be labelled using a \emph{test oracle}~\cite{barr2014oracle}. Therefore, the efficiency of \ddminlocalize depends on both the cost of \ddmin and the performance of the test oracle employed.    
\section{Experimental Setup}

The evaluation of \ddminlocalize was structured around the following research questions.

\subsection{Research Questions}
\begin{enumerate}[leftmargin=*,label=\textbf{RQ.\arabic*}]
    \item\label{rq1} How effectively does \ddminlocalize perform under statement-based, predicate-based and hybrid fault localization?
    \item\label{rq2} How effectively does \ddminlocalize perform under different spectrum-based fault localization techniques?
\end{enumerate}

\subsection{Experimental Subjects}
To evaluate \ddminlocalize and answer the research questions from the previous section, the \emph{Codeflaws}~\cite{Tancodeflaws} and \emph{QuixBugs}~\cite{lin2017quixbugs} were selected based on the following criteria. 

\begin{enumerate}
    \item There should be programs that take \emph{string inputs}. 
    \item There should be a variety of real-world defects that cause \emph{functional bugs / semantic bugs}, i.e., programs producing incorrect or unexpected output for certain inputs.
    \item For each subject, there should be a \emph{golden version}, i.e., a program that produces the expected, correct output for a given input. To obtain the correct label for a test case, the output of the buggy program is compared with the output of the golden version for the same input. If both outputs are different, the label of the test case is considered \emph{failing}; otherwise, it is considered \emph{passing}. Also, the faulty statements of a buggy program are identified by examining the differences between its statements and the corresponding statements in the golden version. 
    \item For each subject, there should be a \emph{manually constructed} test suite.
    \item For each subject, there should be at least one \emph{failing} test case in the manually constructed test suite, i.e., a test input for which the buggy program and the golden program produce different outputs.
\end{enumerate}

The programs in \textbf{Codeflaws}~\cite{Tancodeflaws} have been written in the \textbf{C} programming language. In \textbf{QuixBugs}~\cite{lin2017quixbugs}, the same set of programs has been implemented in both Python and Java programming languages. These two benchmarks have been used in many recent studies on automated program repair~\cite{gazzola2017automatic}. However, to the best of available knowledge, the effectiveness of spectrum-based fault localization on the programs of these benchmarks has not been specifically examined.

Based on the first two criteria, 4 Python programs from QuixBugs and 132 C programs from  Codeflaws were selected for evaluation. Programs taking mixed inputs (e.g. numbers and strings together) and programs exhibiting fleaky behaviour were excluded from the selection. The selected programs contain $11 - 115$ lines.

\subsection{Setup and Evaluation}\label{sec:exp_setup}

In each selected program, \ddminlocalize (Algorithm~\ref{alg:ddmin_loc}) is applied to each failing test case. After \ddmin generates a set of passing and failing test cases, the suspiciousness scores are calculated for each executable statement and predicate separately in the subject program. The hybrid score of a statement is then calculated using Equation~\ref{eqn:hybrid}. 

According to Equation~\ref{eqn:hybrid}, the statement-based approach corresponds to $\alpha=0$, and the predicate-based approach corresponds to $\alpha=1$. In the hybrid method, a statement may be associated with more than one predicate (e.g. an \texttt{if} condition containing multiple conditions). In such cases, the maximum suspiciousness score among the predicates is used for the calculation. 

In the three approaches-statement-based, predicate-based and hybrid, the program elements (statements or predicates) are sorted in descending order of their suspiciousness scores, i.e., thereby producing a ranking of program elements. Here, the statement-based and predicate-based approaches are considered individually as an ablation study of the hybrid method.   

\subsubsection{Determining the faulty statements of a buggy program in the experiments:} Following the works of Zou et al.~\cite{zou2019empirical} and Sohn et al.~\cite{sohn2017fluccs}, the faulty statements of a buggy program are identified by examining the differences between its statements and the corresponding statements in the golden version. Sometimes, new statements may be inserted into the golden version to fix the bug. In such situations, following the work of Pearson et al.~\cite{PearsonICSE}, the executable element immediately after the inserted element is considered the one that should be reported by the fault localization technique. 

In some subjects, there may be multiple faulty lines. In such cases, localizing any faulty element is considered a correct localization of the fault.

\subsubsection{Evaluation metrics}
To evaluate the effectiveness of fault localization, \emph{Exam Score} and \emph{Inspect Score} are used. 

\emph{Exam Score}~\cite{wong2008crosstab} is the percentage of elements that has to be examined until a faulty element is found. Lower scores indicate better performance of the technique. \emph{Inspect Score} or \emph{inspect@n} measures whether the faulty statement is ranked within the top $n$ positions. This work considers \emph{inspect@1}, \emph{inspect@3}, \emph{inspect@5} and \emph{inspect@10} similar to \cite{zou2019empirical}. These two metrics have been used in many works related to fault localization. 

A set of program elements might receive the same suspiciousness score as the faulty element. When this occurs, the mid-point of that program element group is taken as the position at which the faulty element is found in calculating the exam score. In inspect score calculation, the \emph{expected rank} of the faulty statement is calculated using the method proposed in \cite{zou2019empirical}. 

After obtaining a sorted list of suspiciousness scores, the exam score and the inspect score are calculated. In the predicate-based method ($\alpha=1$-Equation~\ref{eqn:hybrid}), the predicates are mapped back to their corresponding lines when computing the suspiciousness scores of the statements. For each subject, exam scores are averaged across all the failing test cases. Also, the number of generated test suites that achieved the expected inspect score is also recorded. These results are used to answer \ref{rq1} and \ref{rq2}.

For the experiments of this work, the following values were fixed.

\begin{itemize}
    \item \emph{Timeout:} For each failing test case in a subject, 15 minutes were allocated for running \ddminlocalize under an SBFL algorithm (Table~\ref{tab:fault_localize_alg}).
    \item $\alpha$ \emph{in the Hybrid approach}: A value of $\alpha=0.5$ was used in the Hybrid approach to ensure equal contributions from the predicate-based method and statement-based method.
\end{itemize}

\section{Experimental Results}\label{sec_exp_res}

\begin{figure}[htbp]
    \centering
    \includegraphics[width=\linewidth]{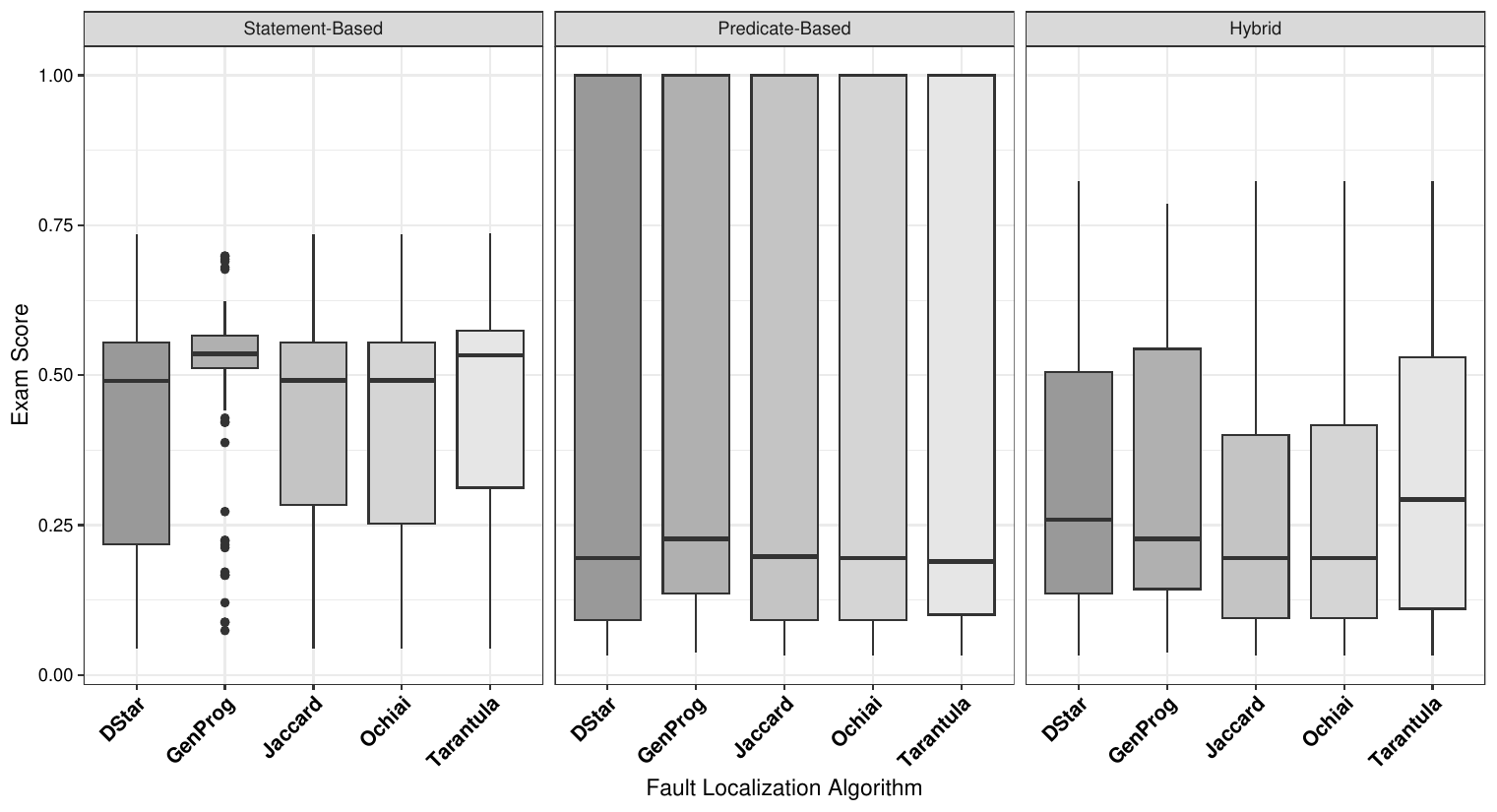}
    \caption{Exam scores under different program elements and different SBFL techniques}
    \label{fig:res_exam_score}
\end{figure}

\begin{figure}[htbp]
    \centering

    \begin{subfigure}{0.5\textwidth}
        \centering
        \includegraphics[width=\linewidth]{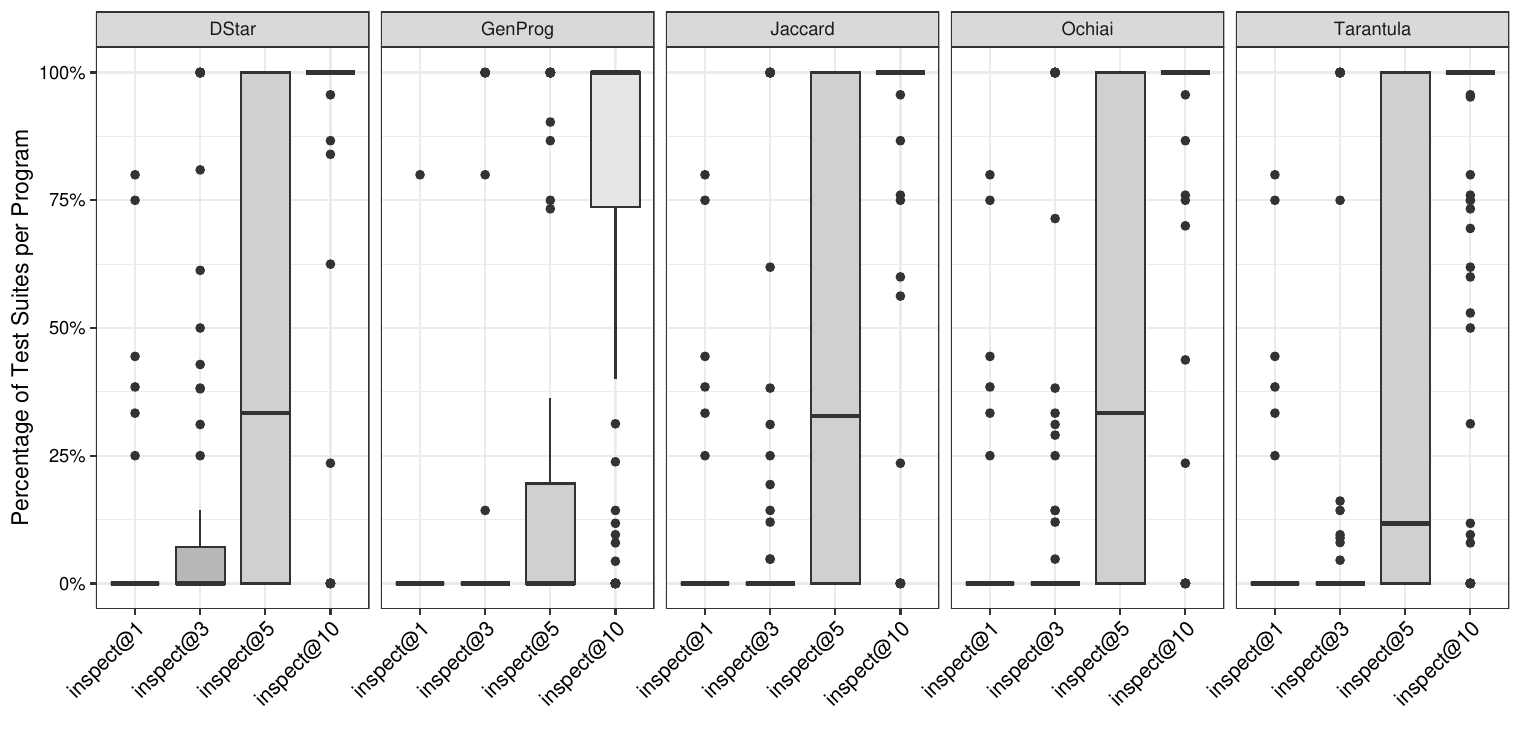}
        \caption{Statement-Based}
        \label{fig:inspect_statement}
    \end{subfigure}
    
    \begin{subfigure}{0.5\textwidth}
        \centering
        \includegraphics[width=\linewidth]{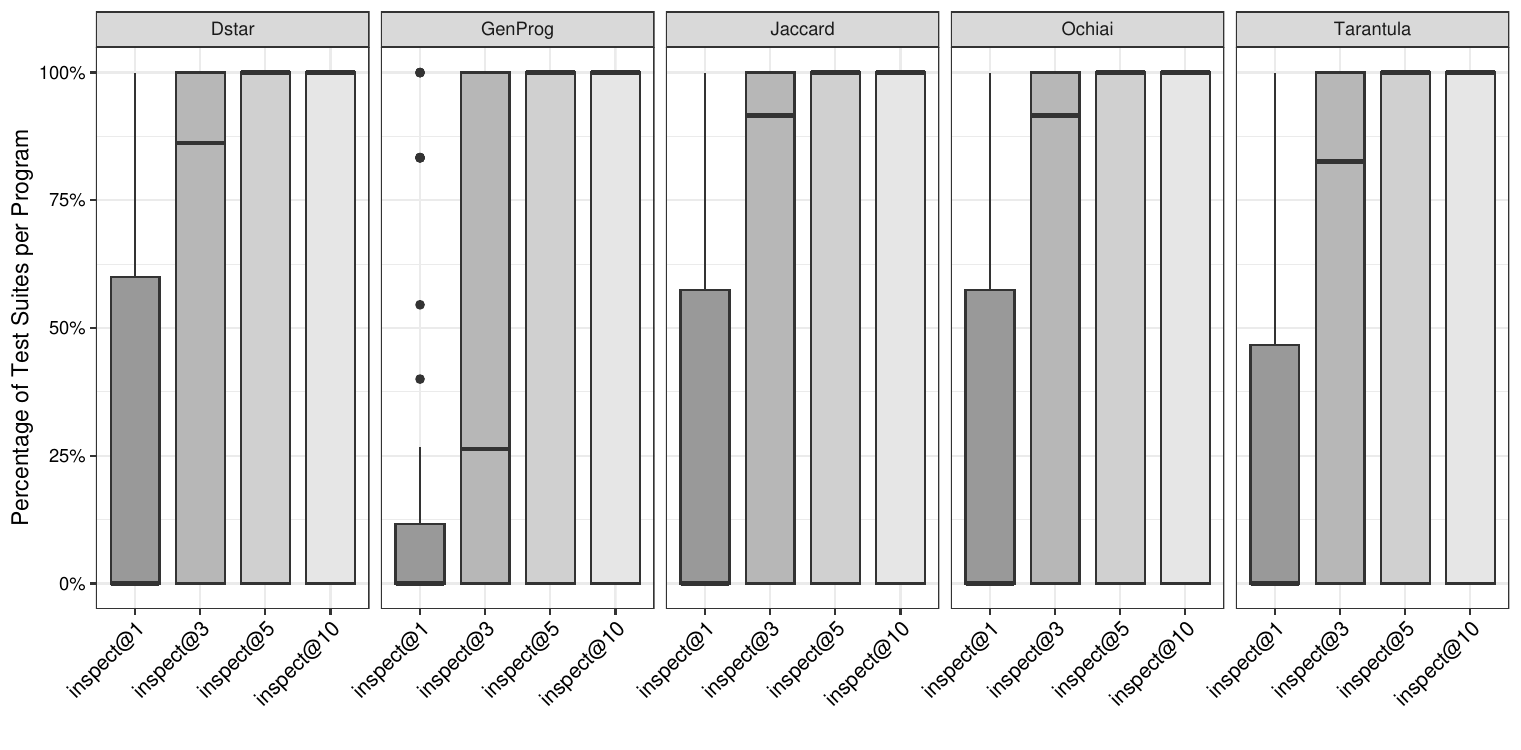}
        \caption{Predicate-Based}
        \label{fig:inspect_predicate}
    \end{subfigure}
    
    \begin{subfigure}{0.5\textwidth}
        \centering
        \includegraphics[width=\linewidth]{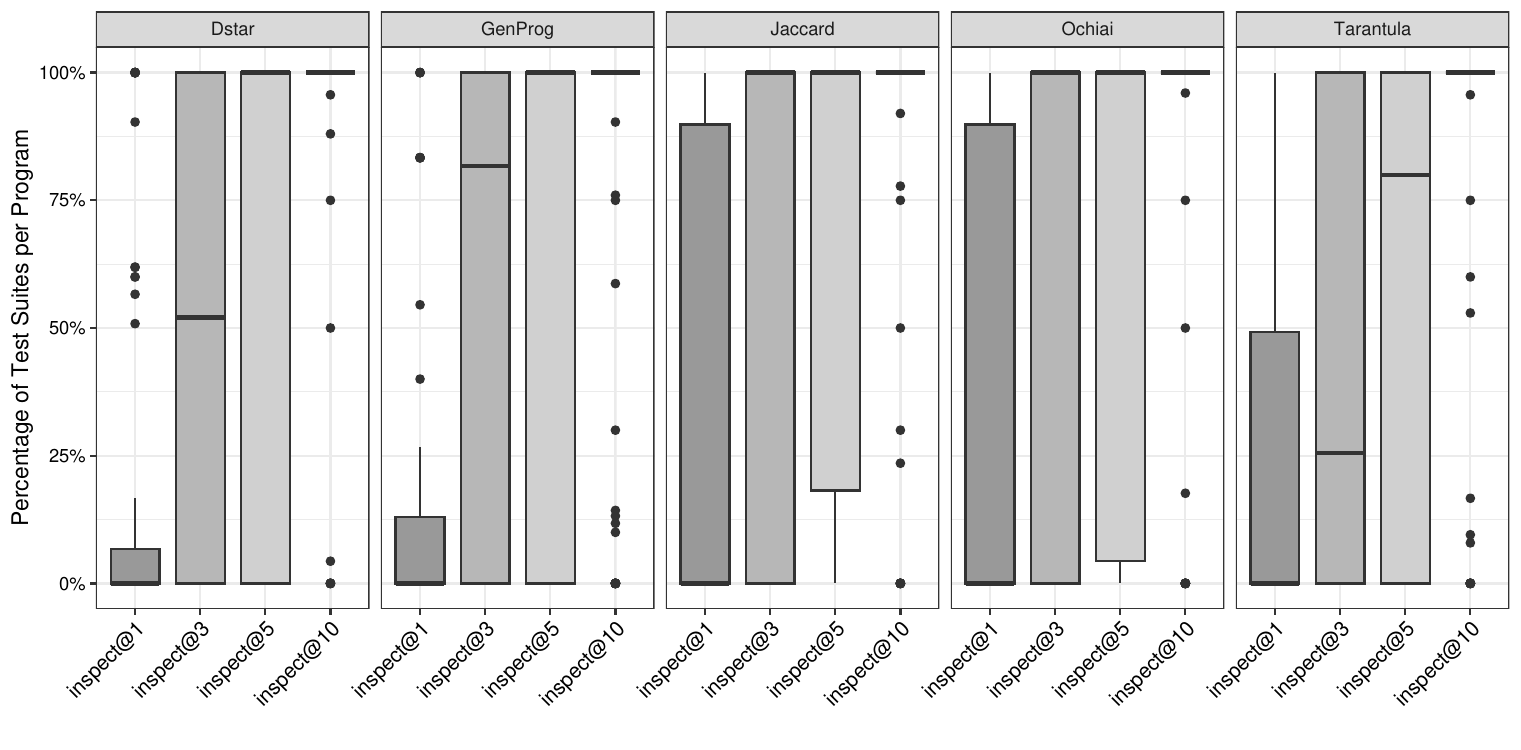}
        \caption{Hybrid}
        \label{fig:inspect_hybrid}
    \end{subfigure}
    
    \caption{Percentage of \ddminlocalize generated test suites per each subject that achieved the inspect scores in Statement-Based, Predicate-Based and Hybrid methods}
    \label{fig:inspect_score}
\end{figure}

Fig.~\ref{fig:res_exam_score} shows the exam scores of \ddminlocalize  with each SBFL algorithm (Table~\ref{tab:fault_localize_alg}) for program elements and the hybrid method (Section~\ref{meth:prog_element}). Fig.~\ref{fig:inspect_score} shows the subject-wise distribution of \emph{inspect@1}, \emph{inspect@3}, \emph{inspect@5} and \emph{inspect@10} for the test suites generated by \ddminlocalize. 

\subsection{RQ.1: Fault Localization Effectiveness under Different Program Elements}\label{sec:res_rq1}

\begin{tcolorbox}
In most subjects, \ddminlocalize ranks the faulty statements within the top 5 positions in all the generated test suites when using the SBFL algorithms \emph{GenProg}, \emph{Jaccard} and \emph{Ochiai} under the \emph{hybrid method}. Moreover, the median exam scores are below 0.25 when using these SBFL algorithms.   
\end{tcolorbox}

\emph{Statement-based approach:} The median exam scores of all the SBFL algorithms are close to or above 0.5. (Fig.~\ref{fig:res_exam_score} - Statement-based plot). Also, the faulty statements of the programs are ranked within the top 10 positions in most subjects (Fig.~\ref{fig:inspect_statement}). 

\emph{Interpretation:} Given that the exam scores approach or exceed 0.5, \ddminlocalize performs at a near-random level in the statement-based method. The Inspect scores of the subjects also confirm this fact. In this setting, the developer would need to inspect roughly half of the executable lines of the programs to locate the fault, which is comparable to the effort required under a random baseline. Therefore, \ddminlocalize is not effective under the statement-approach.

\emph{Predicate-based approach:}  \ddminlocalize shows the \emph{lowest median exam scores} under the SBFL algorithms in the predicate-based approach (Fig~\ref{fig:res_exam_score}- Predicate-based plot). In most subjects, \ddminlocalize ranks the faulty statements within the top 3 positions (Fig.~\ref{fig:inspect_predicate}) in all the generated test suites.  However, the variances of the exam scores are higher than in the other approaches. The third quartiles of all the SBFL algorithms are 1.0. The reason is that some bugs in the selected program subjects are not associated with conditional statements (E.g. Incorrect assignments). 

\emph{Interpretation:} \ddminlocalize performs better in the predicate-based approach than in the statement-based one. However, the predicate-based approach is not stable, as it cannot effectively identify faulty statements that are not associated with conditional statements. 

\emph{Hybrid approach:}  Compared to the predicate-based approach, the variances of the exam scores are lower in the hybrid approach (Fig.~\ref{fig:res_exam_score}- Hybrid-plot). The variances of \emph{inspect@10} are significantly lower in all the SBFL algorithms in the hybrid approach compared to the predicate-based method (Fig.~\ref{fig:inspect_hybrid}). Overall, \ddminlocalize outperforms the statement-based method in the hybrid approach in terms of both exam scores and inspect scores. 

\emph{Interpretation:} The hybrid approach, which combines both statement-based and predicate-based methods, is effective in locating the faulty statements for many program subjects. The limitations in the statement-based and predicate-based approaches have been addressed in the hybrid method. 

\textbf{Result.}~\ddminlocalize \emph{shows its best performance in the hybrid approach} 
\subsection{RQ.2: Fault Localization Effectiveness under Different SBFL techniques}

\begin{tcolorbox}
    \ddminlocalize shows the lowest median scores in the hybrid method when using Jaccard and Ochiai. The median exam scores in this setting are below 0.2. \ddminlocalize. Moreover, in most program subjects, both the SBFL algorithms can rank the faulty statements within the top 3 positions in all the generated test suites.
\end{tcolorbox}

As described in Section~\ref{sec:res_rq1}, none of the SBFL algorithms are effective in locating the faulty statements of the subject programs in the \emph{statement-based approach}. In the \emph{predicate-based} approach, the median exam scores achieved by the SBFL algorithms are below 0.25 and, except for Genprog, these scores are nearly equal. 
(Fig.~\ref{fig:res_exam_score}-Predicate-based plot). The paired \emph{Wilcoxon-test}~\cite{rosner2006wilcoxon} conducted on each pair of SBFL algorithms confirms this observation ($p>0.05$). 

In the hybrid method, the variances of exam scores for all the SBFL algorithms are lower than those in the predicate-based method, for the reason explained in Section~\ref{sec:res_rq1}. Also, the median exam scores of Ochiai and Jaccard are equal ($0.19$) and are the lowest among the algorithms (Fig.~\ref{fig:res_exam_score}-Hybrid plot). However, the paired \emph{Wilcoxon-test}~\cite{rosner2006wilcoxon} indicates that \ddminlocalize performs better with Jaccard than with Ochiai ($p<0.05$). In most program subjects, the faulty statements are ranked within the top 3 positions in all the generated test suites when using these two SBFL algorithms (Fig.~\ref{fig:inspect_hybrid}).  

Another important observation is that the median exam scores of DStar and Tarantula are higher in the hybrid method than in the predicate-based method (Fig.~\ref{fig:res_exam_score}). Moreover, the \emph{inspect@3} values show a decrease in the hybrid method compared to the predicate-based method. This occurs because the characteristics of these SBFL algorithms negatively affect the hybrid score calculation in Equation~\ref{eqn:hybrid}.

\emph{Interpretation:} The SBFL techniques do not perform well with the test cases generated with \ddmin in the statement-based method. These techniques perform better in the predicate-based method; however, those fail to localize the faulty statements in some programs. Compared to the other two methods, the hybrid approach improves the fault localization effectiveness of the SBFL techniques. Among the SBFL techniques, \ddminlocalize shows its best performance when the \emph{Jaccard score} is used.

\textbf{Result.}~\ddminlocalize \emph{effectively localizes faulty statements when the Jaccard score is used with the hybrid approach.} 


\section{Discussion and Future Work}
\balance
\emph{Delta Debugging Minimization} (\ddmin)~\cite{zeller2002simplifying}, proposed by Zeller et al., is an effective algorithm for exploring the minimal failing input exposing a bug. It is particularly useful for test case minimization. Also, it can be used to locate the statements that cause a compiler to crash~\cite{artzi2010directed}. However, \ddmin cannot be used to localize faulty statements in other types of bugs, especially in \emph{semantic bugs}~\cite{tan2014bug}. To address this issue, this paper introduces \ddminlocalize, which combines delta debugging minimization with spectrum-based fault localization. 

According to the results (Section~\ref{sec_exp_res}), \ddminlocalize achieves its best performance when combined with Jaccard score, yielding a median exam score below $0.2$. This performance is better than, or comparable to,  the results reported in recent works by Jones et al.~\cite{jones2005empirical} and Wong et al.~\cite{wong2016survey} and Abreu et al.~\cite{abreu2009spectrum}. Therefore, the performance of \ddminlocalize is consistent with state-of-the-art techniques. In addition, in this setting, the faulty statements are ranked within the top 3 positions in all the generated test suites for most subjects (Fig.\ref{fig:inspect_hybrid}). This result implies that a developer would require relatively little effort to locate the faulty statements with \ddminlocalize.

The results in Section~\ref{sec_exp_res} indicate that the predicate-based approach does not perform well for certain bugs, which results in the higher variance observed in the exam scores (Fig.~\ref{fig:res_exam_score}). These cases correspond to bugs that are not associated with conditional statements (E.g. incorrect value assignment, incorrect arithmetic computations, etc.). This limitation is successfully addressed by the hybrid method. 

In the hybrid method, DStar and Tarantula show higher median scores than in the predicate-based method (Fig.~\ref{fig:res_exam_score}). Their \emph{inspect@3} scores also exhibit lower medians compared to the predicate-based method (Fig.~\ref{fig:res_exam_score}). This behaviour stems from the properties of these SBFL algorithms. DStar~\cite{wong2008crosstab} assigns an infinite score to predicates and statements executed solely by all failing test cases (Table~\ref{tab:fault_localize_alg}). In such cases, the hybrid score (Equation~\ref{eqn:hybrid}) of such a statement remains unchanged, even when no faulty predicate is associated with it. Tarantula~\cite{jones2005empirical} is highly sensitive to the ratio between passing and failing test cases.  For some failing inputs, \ddmin may generate an asymmetric test suite, leading to reduced fault-localization accuracy. This issue with Tarantula is observable in both the statement-based and hybrid methods. The aforementioned issues in both DStar and Tarantula contribute to reduced performance in the hybrid method. 

\ddminlocalize can be considered an automated debugging technique that requires only one failing test case. It leverages test inputs generated by \ddmin while exploring the minimal failing input in combination with spectrum-based fault localization. At present, \ddminlocalize is limited to programs taking string inputs. In future work, new techniques will be explored to extend this concept to other types of program inputs. In addition, integrating other automated test generation techniques, such as fuzzing~\cite{liang2018fuzzing}, with \ddminlocalize will also be investigated. Moreover, exploring methods for incorporating \emph{program slicing}~\cite{xu2005brief} is another direction that will be considered.  

\section{Related Work}

The key idea of this paper was motivated by the work \textsc{GRAMMAR2FIX}~\cite{grammar2fix}. \textsc{GRAMMAR2FIX} also uses the test cases generated by \ddmin to train an automated test oracle for a bug. In this work, those test cases are instead used for fault localization.  In addition, the hybrid score calculation method, which combines statement-level and predicate-level suspiciousness scores, was inspired by the work of Xuan et al.~\cite{xuan2014learning}. 

Delta Debugging Minimization (\ddmin)~\cite{zeller2002simplifying} was used by Gupta et al.~\cite{gupta2005locating} for fault localization. This work integrates \ddmin with \emph{dynamic slicing}~\cite{xu2005brief}. In contrast to \ddminlocalize, their method identifies a failure inducing region of the code. Moreover, it has been tested only on memory related bugs. Nevertheless, the work of Gupta et al.~\cite{gupta2005locating} provides useful insights into the capabilities of \ddmin. 

 The works of Zou et al.~\cite{zou2019empirical}, Jiang et al.~\cite{jiang2019combining}, K{\"u}{\c{c}}{\"u}k et al.~\cite{kuccuk2021improving} and  Pearson et al.~\cite{PearsonICSE} have explored different approaches to improve fault localization.  Jiang et al.~\cite{zou2019empirical} present an empirical study of a wide range of fault localization techniques on real-world faults and further investigate ways to enhance fault localization by combining different families of fault localization techniques. Their conclusions suggest that combined techniques outperform individual techniques.  The works of Jiang et al.~\cite{jiang2019combining} and  K{\"u}{\c{c}}{\"u}k et al.~\cite{kuccuk2021improving} focus on incorporating program predicates for fault localization. The works by Liu et al.~\cite{liu2006failure} and Nainar et al.~\cite{arumuga2007statistical} are some other fault localization approaches that use program predicates. The findings of these studies suggest that program predicates are an effective source for fault localization. 

 Briand et al.~\cite{briand2007using} use machine learning to improve the Tarantula. This work uses \emph{decision trees} to model failure conditions. The work by Zhang et al.~\cite{zhang2019empirical} uses the \emph{PageRank} algorithm to improve spectrum-based fault localization. A common characteristic of these two works is that they require a large number of test cases to achieve effective fault localization. The works of Zheng et al.~\cite{zheng2016fault}, Rafi et al.~\cite{rafi2024towards}, Wong et al.~\cite{wong2011effective} and Maru et al.~\cite{maru2019effective} are examples of \emph{neural network-based} fault localization approaches. In general, these methods require a set of labelled test cases in advance to proceed. Also, these methods focus only on statement-level coverage. In contrast, \ddminlocalize can operate starting from a single failure-inducing input and considers both statement-level and predicate-level details.

\section{Conclusion}

This paper introduces \ddminlocalize for performing fault localization using a single input that exposes a bug. \ddminlocalize extends Delta Debugging Minimization to support spectrum-based fault localization. Experimental results demonstrate that \ddminlocalize performs best with the Jaccard score in the hybrid method. Moreover, using \ddminlocalize, a developer would typically need less effort to locate the faulty lines of a buggy program.  This method can be applied to programs taking string inputs, regardless of size, provided that the source code is accessible and a test oracle exists to identify test failures. As future work, \ddminlocalize can be extended to support additional input types and integrated with automated test generation and program analysis techniques. 

\section*{Acknowledgements}
I would like to express my sincere gratitude to the \emph{Department of Computer Science, Faculty of Science, University of Ruhuna, Sri Lanka}, and to the \emph{University of Colombo School of Computing (UCSC), Sri Lanka,} for supporting me by providing the computational resources required for this research. 

\bibliographystyle{IEEEtran}
\bibliography{references}
\end{document}